\documentclass[prx,reprint,amsmath,amssymb,floatfix,citeautoscript,superscriptaddress]{revtex4-1}
\usepackage{cancel}
\usepackage{amsmath}
\usepackage{physics}
\usepackage{graphicx}
\usepackage{amssymb}
\usepackage{amsthm}
\usepackage{bm}
\usepackage{dcolumn}
\usepackage{braket}
\usepackage{longtable}
\usepackage{tabularx}

\usepackage{ragged2e}
\usepackage{txfonts}
\usepackage[normalem]{ulem}

\usepackage{color}
\usepackage[usenames,dvipsnames]{xcolor}
\definecolor{myblue}{rgb}{0,0,1}
\usepackage[breaklinks=true,colorlinks=true,linkcolor=myblue,urlcolor=myblue,citecolor=myblue]{hyperref}

\let\vu\undefined
\newcommand{\vu}{{\bm{u}}}
\newcommand{\vp}{{\bm{p}}}
\newcommand{\vk}{{\bm{k}}}
\newcommand{\vm}{{\bm{m}}}
\newcommand{\vn}{{\bm{n}}}
\newcommand{\vx}{{\bm{x}}}

\allowdisplaybreaks

\begin{document}

\title{Anharmonic Lattice Dynamics from Vibrational Dynamical Mean-Field Theory}

\author{Petra Shih}
\affiliation{Department of Chemistry,
Columbia University, New York, New York 10027, USA}
\author{Timothy C. Berkelbach}
\email{t.berkelbach@columbia.edu}
\affiliation{Department of Chemistry,
Columbia University, New York, New York 10027, USA}
\affiliation{Center for Computational Quantum Physics, Flatiron Institute, New   York, New York 10010, USA}

\begin{abstract}
We present a vibrational dynamical mean-field theory (VDMFT) of the dynamics of
atoms in solids with anharmonic interactions.  Like other flavors of DMFT, VDMFT
maps the dynamics of a periodic anharmonic lattice of atoms onto those of a
self-consistently defined impurity problem with local anharmonicity and coupling
to a bath of harmonic oscillators.  VDMFT is exact in the harmonic and molecular
limits, nonperturbative, systematically improvable through its cluster
extensions, usable with classical or quantum impurity solvers (depending on the
importance of nuclear quantum effects), and can be combined with existing
low-level diagrammatic theories of anharmonicity.  When tested on models of
anharmonic optical and acoustic phonons, we find that classical VDMFT gives good
agreement with classical molecular dynamics, including the temperature
dependence of phonon frequencies and lifetimes.  Using a quantum impurity
solver, signatures of nuclear quantum effects are observed at low temperatures.
We test the description of nonlocal anharmonicity via cellular VDMFT and the
combination with self-consistent phonon (SCPH) theory, yielding the powerful
SCPH+VDMFT approach.
\end{abstract}

\maketitle

\section{Introduction}

Since the seminal work of the early twentieth century, phonons have been
foundational for the description of solids.  However, early on it was recognized
that anharmonic effects, corresponding to interactions between phonons, were
nonnegligible and responsible for a variety of phenomena including thermal
expansion, the stability of certain phases, the temperature dependence of phonon
frequencies, phonon lifetimes, and thermal
conductivity~\cite{Clyde1971,Klein1972,Maris1977,Fultz2010,Grimvall2012}.  
For example, many of the structural and dynamical properties of halide and oxide
perovskites have been linked to their soft phonon modes and associated strong
anharmonicity~\cite{Yaffe2017,Marronnier2017,Zhou2018,GoldParker2018,Gehrmann2019,Klarbring2020}.
Moreover, recent work has observed strongly correlated phonon behavior, such as
Kondo-like phonon scattering in thermoelectric clathrates~\cite{Ikeda2019} and
the saturation or violation of Planckian bounds on thermal
transport~\cite{Wu2021,Tulipman2021}.

Following on the self-consistent phonon (SCPH)
theory~\cite{Hooton1958,Werthamer1970,Klein1972}, a number of computational
approaches have been developed to simulate the properties of anharmonic
solids~\cite{Souvatzis2008,Souvatzis2009,Hellman2013,Errea2014,Tadano2015,Tadano2018},
most of which are \textit{static} mean-field theories that seek an optimized
harmonic description of anharmonic systems. Therefore, they yield improved
thermodynamic properties and shifts in phonon frequencies, but cannot predict
phonon lifetimes or non-quasiparticle effects.  Such effects can be partially
described using perturbation
theory~\cite{Maradudin1962,Cowley1965,Klemens1966,Menendez1984,Turney2009,Sun2010},
which fails for strong anharmonicity, or by molecular dynamics
(MD)~\cite{Ladd1986,Tretiakov2004,Turney2009,Koker2009,Sun2010}, which is
computationally expensive when accurate forces are used and requires approximate
techniques to include nuclear quantum
effects~\cite{Ceriotti2009,Dammak2009,Rossi2016,Cheng2018}. 

Here, we present a vibrational \textit{dynamical} mean-field theory (VDMFT)
without the above limitations---it is nonperturbative, exact in the harmonic and
molecular limits, systematically improvable through cluster extensions,
applicable to problems with or without nuclear quantum effects, and describes
phonon spectra.  Our VDMFT is completely analogous to conventional
DMFT~\cite{Georges1992,Georges1996,Vollhardt2011}: it is a many-body theory of
the phonon Green's function (GF)~\cite{Cowley1963,Mahan2000} that maps the
dynamics of an anharmonic lattice onto those of a self-consistently defined
impurity problem. In this way, VDMFT treats local anharmonicity
nonperturbatively. Nonlocal anharmonicity can be included at lower-levels of
theory and through cluster extensions of
DMFT~\cite{Hettler2000,Kotliar2001}---here we focus on cellular VDMFT. 

The layout of this article is as follows. In Sec.~\ref{sec:theory}, we present
the general theory of (cellular) VDMFT. In Sec.~\ref{sec:results}, we present
results for two problems. First, we study a model of optical phonons with local
quartic anharmonicity, and we apply single-site VDMFT with both classical and
quantum impurity solvers. Second, we study a model of acoustic phonons arising
from pairwise Lennard-Jones interactions, and we demonstrate the convergence
behavior of cellular VDMFT and the treatment of nonlocal anharmonicity at the
mean-field level with SCPH theory.  In Sec.~\ref{sec:conc}, we conclude by
identifying future directions.

\section{Theory} 
\label{sec:theory}

Within the Born-Oppenheimer approximation, the vibrational lattice Hamiltonian is
\begin{equation}
H = \sum_{\vn\alpha} \frac{\vp_{\vn\alpha}^2}{2m_\alpha} + \mathcal{V}(\{\vx_{\vn\alpha}\})
\end{equation}
where $\vn$ are lattice translation vectors and $\alpha$ indexes atoms in the
unit cell.  We denote the thermal average with respect to this Hamiltonian as 
$\langle \cdots \rangle = \mathrm{Tr}[\cdots e^{-H/k_\mathrm{B}T}]/Z$
where $Z=\mathrm{Tr} e^{-H/k_\mathrm{B}T}$ is the canonical partition function.
Expanding the anharmonic potential energy surface in terms of displacements away from 
the equilibrium lattice positions, $\vu_{\vn\alpha} = \vx_{\vn\alpha}-\langle \vx_{\vn\alpha}\rangle$,
naturally leads to the dynamical matrix
\begin{equation}
\mathcal{D}_{\alpha i,\beta j}(\vk)
    = \frac{1}{\sqrt{m_\alpha m_\beta}} \sum_{(\vm-\vn)} e^{i\vk\cdot(\vm-\vn)} \Phi_{\vm \alpha i,\vn \beta j}
\end{equation}
where 
$\Phi_{\vm \alpha i,\vn \beta j}=\partial^2\mathcal{V}/\partial u_{\vm \alpha i} \partial u_{\vn \beta j}$
is the force constant matrix evaluated at the equilibrium lattice positions and
$i,j$ are Cartesian coordinates.  In this noninteracting limit or in
\textit{static} mean-field theories of anharmonicity, an eigenvalue
problem
\begin{equation}
\label{eq:phonon_eval}
\left[\bm{\mathcal{D}}(\vk) + \bm{\mathcal{W}}(\vk)\right]\mathbf{c}_\lambda(\vk) 
    = \Omega^2_\lambda(\vk) \mathbf{c}_\lambda(\vk),
\end{equation}
defines a set of phonon frequencies $\Omega_\lambda(\vk)$ and collective coordinates
\begin{equation}
u_\lambda(\vk) = N^{-1/2} \sum_{\vn \alpha i} \sqrt{m_\alpha} 
    c_{\vn \alpha i, \lambda}(\vk) e^{-i\vk\cdot \vn} u_{\vn\alpha i}
\end{equation}
for phonon branch $\lambda$, where $\bm{\mathcal{W}}(\vk)$ is the optional
mean-field contribution (see Sec.~\ref{ssec:acoustic} for an example use of SCPH
mean-field theory).  For notational convenience, we define an effective
harmonic matrix in the Cartesian basis
$\mathbf{\Omega}^2(\vk) \equiv \bm{\mathcal{D}}(\vk) + \bm{\mathcal{W}}(\vk)$.

Our object of interest in VDMFT is the finite-temperature phonon GF
\begin{equation}
i\hbar D_{\lambda\lambda^\prime}(\vk,\omega+i\eta) 
    = \int_0^\infty dt e^{i(\omega+i\eta)t} \langle [u_{\lambda}(\vk,t), u_{\lambda^\prime}(-\vk,0)] \rangle;
\end{equation}
henceforth we drop $i\eta$ for notational simplicity and warn that the GF
$\mathbf{D}(\vk,\omega)$ should not be confused with the dynamical matrix
$\bm{\mathcal{D}}(\vk)$.  Defining the corresponding harmonic GF via
$\mathbf{D}_0^{-1}(\vk,\omega) = \omega^2\mathbf{1} - \mathbf{\Omega}^2(\vk)$,
the interacting phonon GF is given by
\begin{equation}
\label{eq:lattice}
\mathbf{D}^{-1}(\vk,\omega) = \mathbf{D}_0^{-1}(\vk,\omega) - 2\mathbf{\Omega}(\vk)\bm{\pi}(\vk,\omega)
\end{equation}
where $\bm{\pi}(\vk,\omega)$ is the phonon self-energy.

Standard perturbative approaches evaluate the self-energy 
$\bm{\pi}(\vk,\omega)$ up to second order in the anharmonicity using cubic and/or quartic
anharmonic force constants of the potential
$\mathcal{V}$~\cite{Tadano2015,Tadano2018}.  Instead, in VMDFT we neglect the
momentum dependence of the self-energy term
$\mathbf{\Omega}(\vk)\bm{\pi}(\vk,\omega) \approx \mathbf{\Omega}\bm{\pi}(\omega)$, which 
is obtained nonperturbatively from a self-consistently defined impurity problem
(the impurity frequency matrix $\mathbf{\Omega}$ will be defined below).  The
Hamiltonian of the impurity problem is of the Caldeira-Leggett
form~\cite{Caldeira1983},
$H_\mathrm{imp} = H_\mathrm{s} + H_\mathrm{b} + H_\mathrm{sb}$ with
\begin{subequations}
\label{eq:Himp}
\begin{align}
\label{eq:ham_sys}
H_\mathrm{s} &= \sum_\alpha \frac{\vp_\alpha^2}{2m_\alpha} + V_\mathrm{loc}(\{\vu_\alpha\}) \\
H_\mathrm{b} &= \frac{1}{2} \sum_{m} \left(p_{m}^2 + \omega_{m}^2 x_{m}^2\right) \\
H_\mathrm{sb} &= \sum_{\alpha i m} c_{\alpha i,m} u_{\alpha i} x_{m},
\end{align}
\end{subequations}
where $(p_{m}, x_{m})$ are degrees of freedom of a bath of harmonic oscillators
and the local potential $V_\mathrm{loc}$ includes {bare} harmonic and anharmonic
interactions within the cell and {the local, harmonic parts of the nonlocal
interactions across the cell boundary (possibly at the mean-field level)}.  The
harmonic bath is completely specified by the hybridization function
$\bm{\Delta}(\omega)$, which captures the influence of the lattice on the
dynamics of the cluster and is defined by 
\begin{equation}
\label{eq:hybridization}
2\mathbf{\Omega} \bm{\Delta}(\omega) = \omega^2\mathbf{1} - \mathbf{\Omega}^2 
    - 2\mathbf{\Omega} \bm{\pi}(\omega) - \mathbf{D}_\mathcal{C}^{-1}(\omega),
\end{equation}
where $\mathbf{D}_\mathcal{C}(\omega) = N^{-1}\sum_\vk \mathbf{D}(\vk,\omega)$
is the cellular GF and $N$ is the number of cells in the Born-von Karman
supercell.  The effective dynamical matrix of the impurity $\mathbf{\Omega}^2$
is determined by the harmonic part of $V_\mathrm{loc}$.  The impurity
Hamiltonian (\ref{eq:Himp}) is related to the hybridization by the spectral
density
$\mathbf{J}(\omega) = -2\mathrm{Im}\mathbf{\Omega}\bm{\Delta}(\omega)$ or
\begin{equation}
J_{\alpha i,\beta j}(\omega) = \frac{\pi}{2} \sum_m \frac{c_{\alpha i,m} c_{\beta j,m}}{\omega_{m}}
    \left[\delta(\omega - \omega_{m}) - \delta(\omega + \omega_{m})\right].
\end{equation}
By construction, the hybridization is exact when the lattice and impurity
problems are treated at the same level of theory (for example, in the harmonic
limit, the bath construction is a simple normal mode transformation of the
lattice degrees of freedom). The power of DMFT lies in the fact that an
accurate treatment of the dynamics of the impurity problem~(\ref{eq:Himp}) is
far more tractable than that of the anharmonic lattice problem. In this
case, where the lattice and impurity problems are treated at different levels of
theory, a self-consistent solution must be obtained.

Various impurity solvers, discussed more below, can be used to calculate the
anharmonic impurity GF and phonon self-energy,
\begin{align}
i\hbar[\mathbf{D}_{\mathrm{imp}}(t)]_{\alpha i,\beta j} &= \theta(t) \langle [u_{\alpha i}(t), u_{\beta j}(0) ]\rangle, \\
2\mathbf{\Omega} \bm{\pi}(\omega) 
    &= \mathbf{d}_{\mathrm{imp}}^{-1}(\omega) - \mathbf{D}_{\mathrm{imp}}^{-1}(\omega),
\end{align}
where $\mathbf{d}_\mathrm{imp}^{-1}(\omega) = \omega^2\mathbf{1}-\bm{\Omega}^2-2\bm{\Omega}\bm{\Delta}(\omega)$ 
is the harmonic impurity GF.  Within the DMFT approximation, this phonon
self-energy defines the lattice GF $\mathbf{D}(\vk,\omega)$ and thus the
cellular GF, the hybridization via Eq.~(\ref{eq:hybridization}), and the
impurity problem itself. This establishes the VDMFT self-consistency condition
$\mathbf{D}_\mathcal{C}(\omega) = \mathbf{D}_{\mathrm{imp}}(\omega)$.
In practice, we make an initial guess of the self-energy and iterate the VDMFT
loop until convergence.  Note that with a straightforward redefinition of the
size of a unit cell, the above equations also describe the cellular VDMFT
approach for including short-range nonlocal anharmonicity exactly.  Because
cellular DMFT breaks translational symmetry (beyond a single-cell cluster), we
periodize the converged self-energy to study lattice
quantities~\cite{Kotliar2001,Parcollet2004,Civelli2005}, as discussed below.

Because phonons formally obey Bose-Einstein statistics, our VDMFT has many
similarities to bosonic DMFT~\cite{Hu2009,Anders2011}.  However, the number of
phonons in solids is not conserved, unlike interacting lattice bosons such as
cold atoms.  Thus, equilibrium condensation is not a primary concern, unlike in
most applications of bosonic DMFT.  Note that if the equilibrium atomic
positions are approximated by their zero-temperature values that minimize the
potential energy $\langle \vx_{\vn\alpha}\rangle \approx \vx_{\vn\alpha}^{(0)}$,
then a structural phase transition occurring at elevated temperature is
consistent with $\langle \vu_{\vn\alpha}\rangle \neq 0$, which is sometimes
described as condensation.  In such cases, an explicit treatment of condensation
within VDMFT might be worthwhile. However, this concern is removed as long as
the equilibrium positions are properly redefined~\cite{Yukalov2012}.  We note
that the equilibrium atomic positions are a \textit{static} property, which is
far easier to calculate than a dynamical one, such as the Green's function.  In
principle, the equilibrium atomic positions can be self-consistently defined as
those that minimize the VDMFT vibrational free energy or that of simpler
theories such as SCPH. Moreover, most vibrational problems in solids can be
treated without regard for particle statistics (classically or quantum
mechanically), and these are the target problems for VDMFT.

\section{Results}
\label{sec:results}

\subsection{Optical phonons: Single-site VDMFT}
\label{ssec:optical}

To illustrate VDMFT, we first consider a one-dimensional chain of oscillators
with mass $m=1$, periodic boundary conditions, and \textit{purely local}
anharmonicity,
\begin{equation}
\label{eq:ham_optical}
H = \sum_{n=1}^{N} \left[\frac{p_n^2}{2} + \frac{1}{2}\Omega_0^2 u_n^2 + gu_n^4\right] 
    + \frac{1}{2} \omega_0^2 \sum_{n=1}^{N} (u_n - u_{n+1})^2.
\end{equation}
For the quartic anharmonicity considered here, all frequency shifts and
lifetimes are due to four-phonon processes or higher.  Physically, this
Hamiltonian could model a molecular crystal with anharmonic intramolecular
vibrations and harmonic intermolecular vibrations.  Here and throughout we
assume a fixed volume, such that there is no thermal expansion. We emphasize
that, given the mean-field nature of VDMFT, the
one-dimensional models studied here provide a challenging test.  The optical
phonons of the Hamiltonian~(\ref{eq:ham_optical}) are $u(k) = N^{-1/2} \sum_n
e^{-ikn} u_n$ with the noninteracting harmonic dispersion
$\Omega^2(k) = \Omega_0^2 + 4\omega_0^2\sin^2(k/2)$. 
Within single-site VDMFT, the impurity Hamiltonian has components
\begin{subequations}
\label{eq:Himp_rubin}
\begin{align}
H_\mathrm{s} &= \frac{p^2}{2} + \frac{1}{2}\Omega_0^2 u^2 + gu^4 + \omega_0^2 u^2 
    \equiv \frac{p^2}{2} + V_\mathrm{loc}(u) \\ 
H_\mathrm{b} &= \tfrac{1}{2} \sum_m \left(p_m^2 + \omega_m^2 x_m^2\right) \\
H_\mathrm{sb} &= u \sum_m c_m x_m,
\end{align}
\end{subequations}
where $J(\omega>0) = (\pi/2) \sum_m c_m^2/\omega_m\delta(\omega - \omega_m)$. 
Note that the local potential includes a harmonic term $\omega_0^2 u^2$ 
arising from the nonlocal interaction across the cell boundary
such that $\Omega = [\Omega_0^2 + 2\omega_0^2]^{1/2}$ is the harmonic impurity frequency.

We first assess the performance of VDMFT with a \textit{classical} impurity
solver.  In this classical limit, the dynamics of the harmonic bath can be
integrated out such that the impurity position satisfies the generalized
Langevin equation (GLE),
\begin{equation}
\label{eq:gle}
\ddot{u}(t) = -\frac{dV_\mathrm{eff}}{du} - \int_0^t ds \gamma(t-s) \dot{u}(s) + \xi(t)
\end{equation}
where 
$\gamma(t) = (2/\pi) \int_0^\infty d\omega \cos(\omega t) J(\omega)/\omega$ 
is a memory kernel,
$V_\mathrm{eff}(u) = V_\mathrm{loc}(u) - \gamma(t=0)u^2/2$ is the local potential
with a bath-induced renormalization~\cite{Weiss2012},
and $\xi(t)$ is a random force satisfying detailed balance
$\langle \xi(t)\xi(s) \rangle = k_\mathrm{B}T\gamma(t-s)$.
In this formulation, the lattice hybridization can be seen to play the role of a
very specific colored-noise thermostat~\cite{Ceriotti2009,Ceriotti2010}.  As
described in App.~\ref{app:classical_imp}, we solve the GLE
numerically~\cite{Tuckerman1993,Ceriotti2010} to yield an ensemble of
trajectories from which we calculate the classical one-sided impurity
autocorrelation function $C_\mathrm{cl}(t) = \langle u(t)u(0)\rangle$. The
impurity GF is then calculated as 
\begin{equation}
D_\mathrm{imp}(t) = -\frac{1}{\hbar\pi}\theta(t)
    \int_{-\infty}^{\infty}d\omega \sin(\omega t) C_\mathrm{cl}(\omega) Q(\omega,T)
\end{equation}
where $Q(\omega,T) = (\hbar\omega/k_\mathrm{B}T)(1-e^{-\hbar\omega/k_\mathrm{B}T})^{-1}$
is a temperature-dependent quantum correction factor that makes
$D_\mathrm{imp}(t)$ exact in the harmonic limit~\cite{Bader1994}.

\begin{figure}[t]
    \centering
    \includegraphics[scale=0.7]{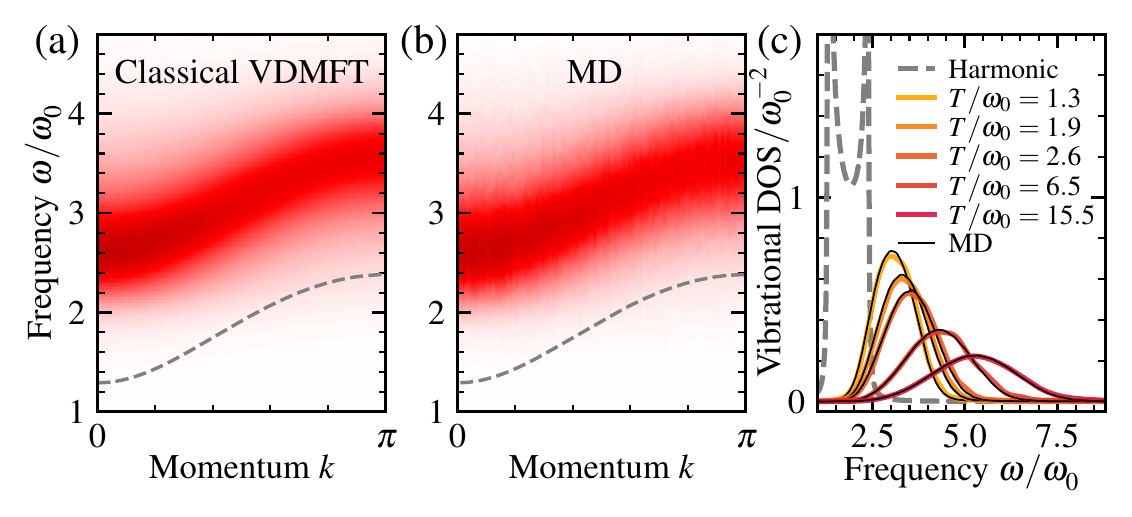}
    \caption{Single-site VDMFT results for the
Hamiltonian~(\ref{eq:ham_optical}) with a classical impurity solver.
(a),(b) Spectral functions from VDMFT and MD, respectively, at
$T/\omega_0=1.3$, compared to the harmonic dispersion (dashed grey line).
(c) Temperature-dependent density of states (DOS) obtained from harmonic theory
(dashed grey), VDMFT (yellow to red), and ``exact'' molecular dynamics (MD, thin solid
black).
In all results, $\eta/\omega_0=0.02$.
}
    \label{fig:optical_classical}
\end{figure}

For the Hamiltonian~(\ref{eq:ham_optical}), we take the harmonic frequency of
the intercellular potential $\omega_0$ as the unit of energy and set
$\hbar=k_\mathrm{B}=1$. We use a local harmonic frequency 
$\Omega_0/\omega_0 = 1.3$ and anharmonicity $g/\omega_0^3 = 4.3$. 
In Figs.~\ref{fig:optical_classical}(a) and (b), we show the converged
momentum-resolved spectral function 
$A(k,\omega) = -\pi^{-1} \mathrm{Im} D(k,\omega)$ at $T/\omega_0=1.3$ 
obtained from VDMFT (a) and from exact MD simulations of the full lattice
problem (b)~\footnote{MD simulations were performed with periodic lattices of
100-200 sites and time correlation functions were calculated by averaging over
an ensemble of up to 600,000 trajectories with initial conditions generated by
Metropolis Monte Carlo.}; the agreement is excellent. (At this low
temperature, nuclear quantum effects are significant---see below---but we can
still assess the accuracy of VDMFT within the consistent approximation of
classical dynamics.) For these parameters, the VDMFT loop converged in about
four iterations when initialized by neglecting the self-energy.  As expected,
the peaks of the spectral functions are significantly shifted from the harmonic
dispersion and are broadened due to phonon lifetime effects.  In
Fig.~\ref{fig:optical_classical}(c), we show the total vibrational density of
states (DOS), $N^{-1}\sum_k A(k,\omega)$, at increasing temperatures ranging from 
$T/\omega_0 = 1.3$ to $T/\omega_0=15.5$. As is well known, the harmonic DOS is
independent of temperature.  The agreement between VDMFT and MD is seen to be
excellent at all temperatures and the DOS shows decreasing lifetimes and phonon
hardening with increasing temperature, as expected for a potential with quartic
anharmonicity.  The remarkable accuracy of single-site VDMFT for this problem
can be largely attributed to the purely local form of the anharmonicity.
Importantly, the computational cost of solving the GLE~(\ref{eq:gle}) for a
single degree of freedom is significantly less than that of MD for the system
sizes needed to obtain converged results.

\begin{figure}[t]
    \centering
    \includegraphics[scale=0.7]{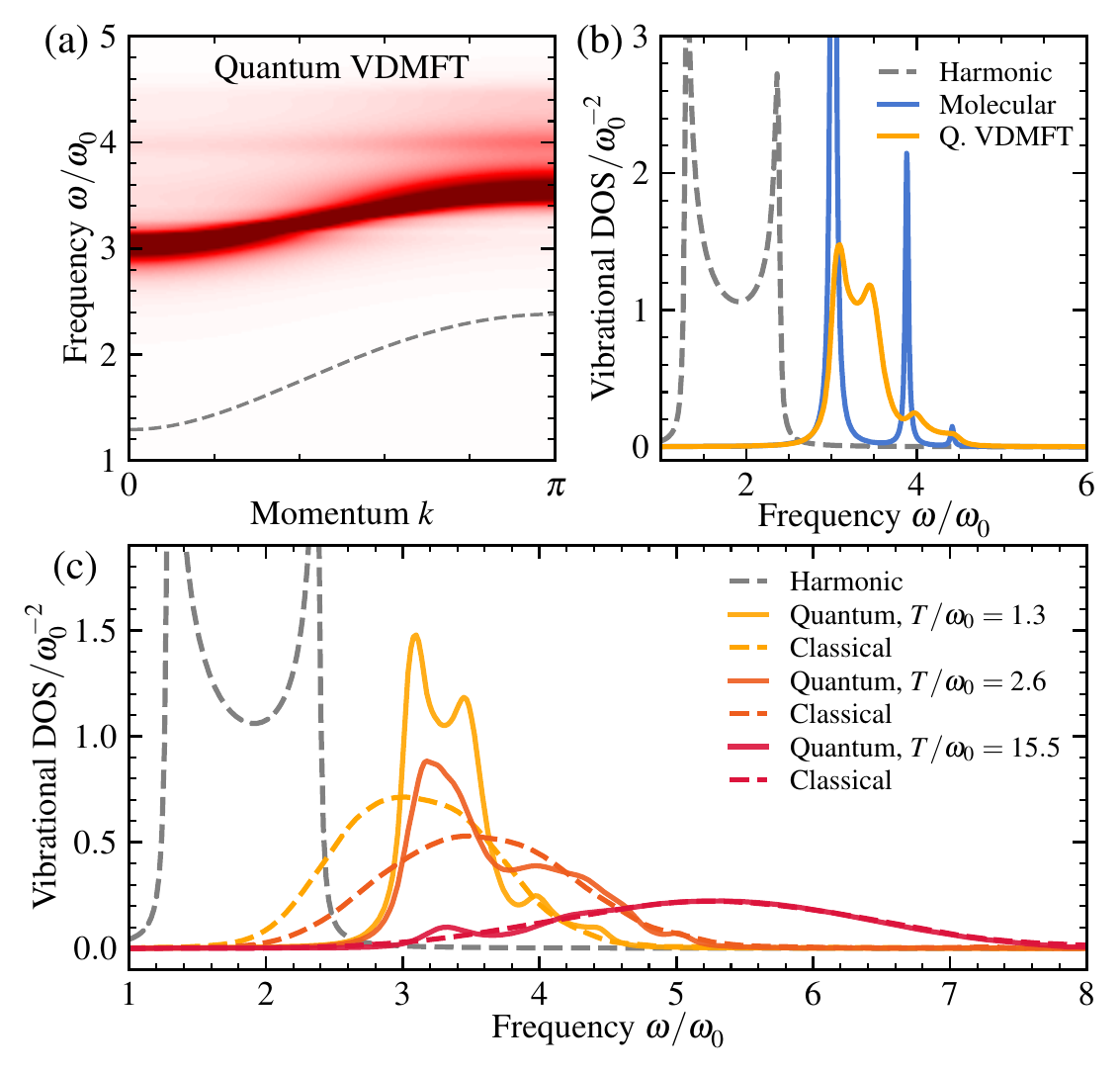}
    \caption{Single-site VDMFT results for the Hamiltonian~(\ref{eq:ham_optical}) with a
quantum impurity solver. (a) Quantum spectral function for the same temperature
as in Figs.~\ref{fig:optical_classical}(a),(b). 
(b) At the same temperature, the DOS from harmonic theory (dashed grey), 
the molecular limit of a single anharmonic oscillator (blue), 
and quantum VDMFT (yellow).
(c) DOS at increasing temperature (yellow to red) obtained by VDMFT with
a quantum (solid) and classical (dashed) impurity solver.
In all results, $\eta/\omega_0 = 0.02$.
}
    \label{fig:optical_quantum}
\end{figure}

Next, we consider the possible importance of nuclear quantum effects, which can
be straightforwardly included in VDMFT with a \textit{quantum} impurity solver.
Here, we use the hierarchical equations of
motion~\cite{Tanimura1989,Ishizaki2005}, which is a numerically exact technique
for simulating the dynamics of systems coupled to harmonic baths; more details
are given in App.~\ref{app:quantum_imp}.  In Fig.~\ref{fig:optical_quantum}(a),
we show the spectral function for the same parameters as in
Figs.~\ref{fig:optical_classical}(a),(b); unlike in the classical case, the
quantum case is a large many-body problem without a numerically tractable exact
solution.  We see that the quantum spectral function is narrower and more
structured than the classical one, indicating that nuclear quantum effects are
indeed important at this relatively low temperature $T/\omega_0 = 1.3$.
Accurate quantum vibrational spectra of a condensed-phase system are
extremely hard to obtain by other means.

To understand the origin of the structured spectral features, in
Fig.~\ref{fig:optical_quantum}(b), we compare the lattice DOS to the analogous
quantum spectrum for a single anharmonic site (the so-called atomic or molecular
limit) with potential $V(u) = \frac{1}{2}\Omega_0^2u^2 + gu^4$,
\begin{equation}
-\pi^{-1}\mathrm{Im}\ d(\omega) = \sum_{ab} (P_a-P_b) |\langle \psi_a | u | \psi_b\rangle|^2 \delta\big(\omega-(E_b-E_a)\big),
\end{equation}
where $|\psi_a\rangle$, $E_a$ are eigenstates and eigenvalues of the anharmonic oscillator
and $P_a = e^{-E_a/T}/\sum_b e^{-E_b/T}$.  The peaks in the molecular spectrum
are thus due to transitions between eigenstates of the anharmonic oscillator
with intensities depending on their Boltzmann weights and transition matrix
elements.  These discrete quantum transitions are responsible for the structure
seen in the lattice DOS when a quantum impurity solver is used.  In
Fig.~\ref{fig:optical_quantum}(c), we compare the quantum and classical DOS at
three temperatures spanning the same range as in
Fig.~\ref{fig:optical_classical}(c). At low temperatures, we see the discrepancy
due to the importance of nuclear quantum effects. However, at high temperatures,
we see that the quantum and classical spectral functions agree due to the
diminishing importance of nuclear quantum effects.

\subsection{Acoustic phonons: Cellular VDMFT}
\label{ssec:acoustic}

Consider now the vibrational Hamiltonian with \textit{nonlocal} anharmonicity due to a pair
potential,
\begin{equation}
\label{eq:ham_acoustic}
H = \sum_{n=1}^{N} \left[\frac{p_n^2}{2} + V(u_{n}-u_{n+1}) \right] 
\end{equation}
Due to its invariance to infinitesimal translations, the above
Hamiltonian will exhibit a single acoustic phonon branch; the noninteracting
harmonic dispersion is $\Omega(k) = 2\omega_0 \lvert\sin(k/2)\rvert$ where
$\omega_0$ is the harmonic frequency of the pair potential $V$.

Treating the nonlocal interactions encoded in a pair potential requires a
cluster VDMFT, such as the cellular extension described above.  In the impurity
Hamiltonian of cellular VDMFT, we keep only the local, harmonic parts of the
nonlocal interactions that cross the boundary of the cluster.  In principle, we
could also keep local, anharmonic parts of these interactions.  However, doing
so breaks the symmetry associated with infinitesimal translations and
incorrectly opens a gap at the $\Gamma$ point (note that periodization only
restores \textit{lattice} translational symmetry).
Here, we study the convergence of cellular VDMFT with cluster size $N_\mathrm{c}$,
using classical impurity solvers, ranging from $N_\mathrm{c}=2\text{--}4$.
After convergence of the DMFT cycle, we calculate the momentum-resolved spectral
function using a periodized self-energy term~\cite{Kotliar2001,Parcollet2004,Civelli2005},
\begin{equation}
\Omega(k)\pi(k,\omega) = N_\mathrm{c}^{-1} \sum_{\alpha\beta} [\bm{\Omega\pi}(\omega)]_{\alpha\beta} e^{ik(\alpha-\beta)},
\end{equation}
although other choices are possible~\cite{Stanescu2006,Sakai2012}.

\begin{figure}[b]
    \centering
    \includegraphics[scale=0.7]{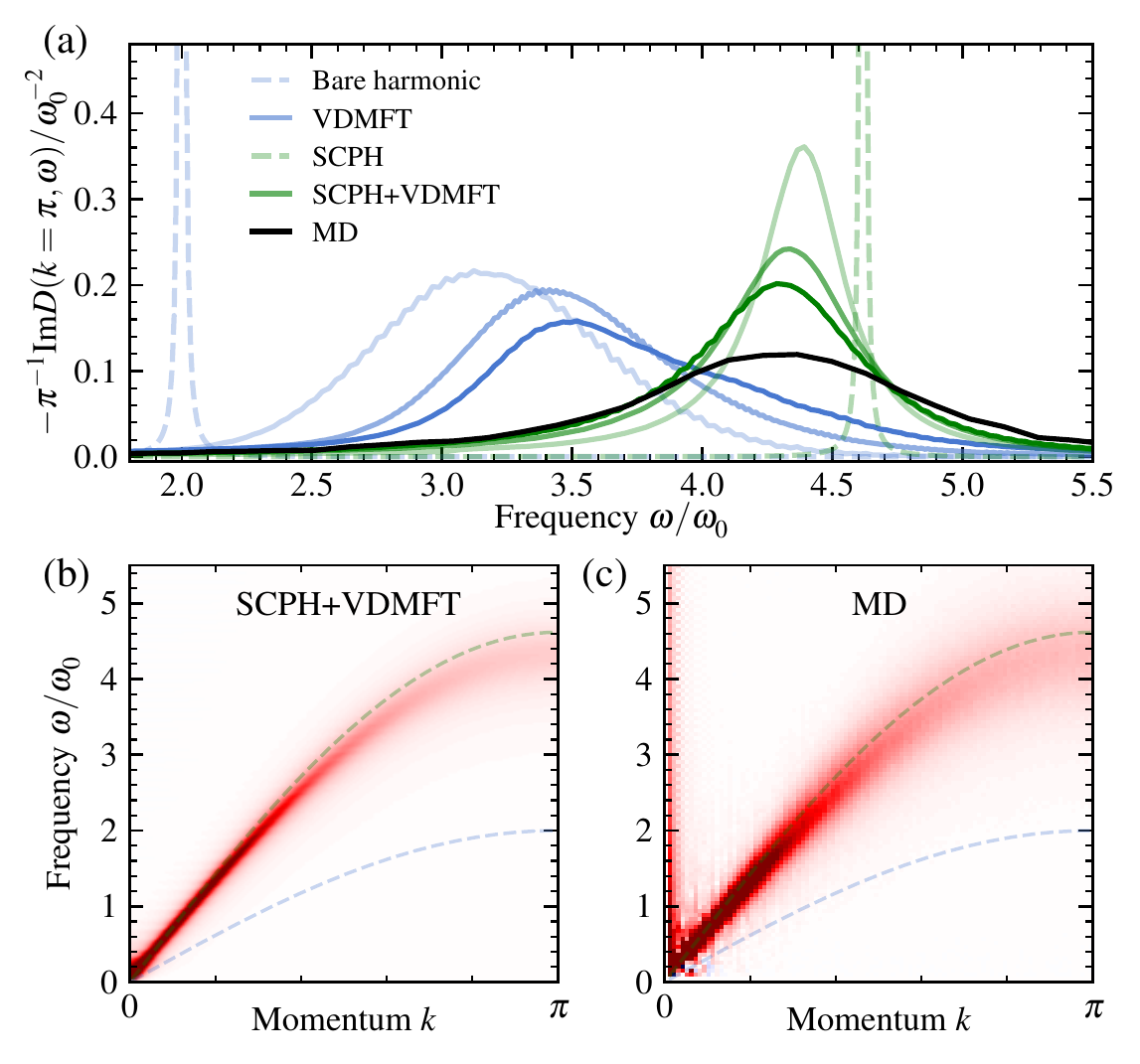}
    \caption{Cellular VDMFT results for the Hamiltonian~(\ref{eq:ham_acoustic})
with a Lennard-Jones potential truncated at fourth order at $T/\omega_0=2.7$.
(a) Spectral function at $k=\pi$ obtained from the bare harmonic theory (dashed
light blue), SCPH theory (dashed light green), and their combinations with
cellular VDMFT as a function of cluster size ($N_\mathrm{c}=2,3,4$, light to
dark), compared to the exact MD result (solid black).  (b) Spectral function
from SCPH+VDMFT with $N_\mathrm{c}=4$ and (c) from MD.  In all results,
$\eta/\omega_0 = 0.01$.
}
    \label{fig:acoustic}
\end{figure}

Results of cellular VDMFT are shown in Fig.~\ref{fig:acoustic} using a
Lennard-Jones pair potential with its minimum at the lattice spacing and its
harmonic frequency $\omega_0$ taken as the unit of energy.  For simplicity, the
potential is truncated to include only cubic and quartic anharmonicity.  Results
are shown at temperature $T/\omega_0=2.7$, where we do not expect significant
nuclear quantum effects.  In Fig.~\ref{fig:acoustic}(a), we show the spectral
function at the zone boundary $k=\pi$.  We see that as the cluster size
increases, cellular VDMFT yields a spectral function that approaches the exact
one from MD.
However, the convergence is slow because of the poor accuracy of the bare harmonic GF
$D_0(k,\omega)$, which completely neglects nonlocal anharmonicity. 

\begin{figure}[t]
    \centering
    \includegraphics[scale=0.7]{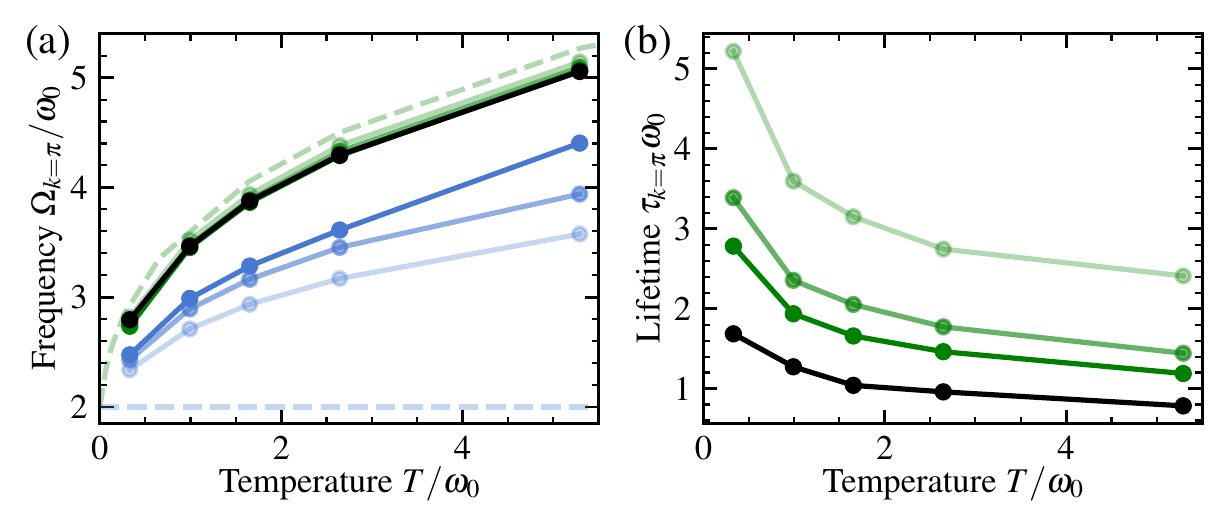}
    \caption{
(a) Temperature-dependent phonon frequency $\Omega_k$ and (b) phonon lifetime
$\tau_k$ at the Brillouin zone boundary $k=\pi$ for the same Lennard-Jones
system as in Fig.~\ref{fig:acoustic}. Colors and linetypes are the same as in
Fig.~\ref{fig:acoustic}.
}
    \label{fig:acousticT}
\end{figure}

For better performance, we can treat nonlocal anharmonicity with an
improved low-level GF such as the one from SCPH theory (i.e., mean-field theory
in the anharmonicity).  For this simple model, the SCPH self-energy is
diagonal.  Considering only cubic and quartic anharmonicity, only the latter
contributes to first order in a loop diagram and the SCPH phonon frequencies
$\Omega(k)$ are determined self-consistently according to
\begin{subequations}
\begin{align}
\Omega^2(k) &= \omega^2(k) + \mathcal{W}(k) \equiv \omega^2(k) + 2\Omega(k) \pi^{(0)}(k) \\
\pi^{(0)}(k) &= k_\mathrm{B}T \sum_{k'} \frac{\Phi(k,-k,k',-k')}{4\Omega(k) \Omega^2(k')},
\end{align}
\end{subequations}
where 
$\mathcal{W}(k)$ is the static mean-field contribution introduced in Eq.~(\ref{eq:phonon_eval})
and
\begin{equation}
\Phi(k,-k,k',-k') = N^{-1} \sum_{n_2,n_3,n_4} \frac{\partial^4 \mathcal{V}}{\partial u_0 \partial u_{n_2} \partial u_{n_3} \partial u_{n_4}}
    e^{-ikn_2+ik'(n_3-n_4)}
\end{equation}
is the quartic force constant.  In the above, we have taken the high-temperature
limit of classical statistics and relabeled the noninteracting phonon
frequencies as $\omega(k)$ to distinguish them from the SCPH frequencies
$\Omega(k)$.  Using the SCPH GF in place of the noninteracting GF in the
VDMFT equations defines the SCPH+VDMFT method. In this way, the SCPH+VDMFT
method treats local anharmonicity exactly and nonlocal quartic anharmonicity at
the mean-field level, analogous to the use of Hartree-Fock+DMFT for
fermionic problems.  Note that the diagrammatic formulation of SCPH theory
ensures a rigorous combination with VDMFT without the need for double-counting
corrections~\cite{Haule2015,Karolak2010}.

The SCPH+VDMFT spectral function at the zone boundary is shown in
Fig.~\ref{fig:acoustic}(a) with increasing cluster size. Because of the improved
performance of SCPH theory, the convergence with cluster size is significantly
improved.  In Fig.~\ref{fig:acousticT}, we study this performance as a function
of temperature, plotting the phonon frequency (a) and lifetime (b) obtained by
the harmonic theory (with no temperature dependence), SCPH theory (with infinite
lifetime, characteristic of static mean-field theories), and the two
flavors of VDMFT, compared to exact results from MD; all lifetimes were
determined by a fit to a Lorentzian lineshape function. 
The exact MD lifetimes $\tau_{k=\pi}\omega_0\approx 1$ are indicative of nearly
incoherent phonon dynamics that are beyond the limits of perturbation theory.
Clearly, SCPH+VDMFT is a significant improvement for the phonon frequency, and
the cellular results converge quickly to the exact MD result. The SCPH+VDMFT
lifetime is not yet converged, and the value for our largest cluster size
exhibits an error of about 50\%.  We do not consider the harmonic+VDMFT lifetime
because the lineshape is significantly asymmetric, as can be seen in
Fig.~\ref{fig:acoustic}(a).

\section{Conclusions and future work}
\label{sec:conc}

We have introduced vibrational DMFT, including a cellular extension {and the
combination with approximate low-level theories, which demonstrates that VDMFT
is not a replacement for existing theories of anharmonicity, but rather a
formalism that enables their systematic and nonperturbative improvement.} Future
work will test alternative impurity solvers and cluster
methods~\cite{Hettler2000}, as well as the performance in higher dimensions and
for other observables such as the free energy~\cite{Sham1965,Plakida1978} and
thermal conductivities~\cite{Pang2013,Liao2015,Zhou2018,GoldParker2018}.
Moreover, VDMFT could be adapted for use in problems with coupling between
electronic and bosonic degrees of freedom, such as those with physical
electron-phonon coupling~\cite{Koker2009} or those arising in extended DMFT to
treat nonlocal interactions~\cite{Pankov2002,Akerlund2014}.

VDMFT can be extended to atomistic materials, either with model
Hamiltonians~\cite{Ai2014,Chen2014} or in a fully ab initio framework. In the
latter case, force-fields or electronic structure theory can be used to
determine the anharmonic potential energy surface; in particular, the small size
of a unit cell will enable the use of highly accurate electronic structure
methods that would otherwise be too costly for explicit MD of large supercells.
The anharmonic impurity problem can then be solved using thermostatted
MD~\cite{Ceriotti2009,Ceriotti2010} or efficient quantum configuration
interaction approaches~\cite{Neff2009,Fetherolf2021}.  Work along all of these
lines is currently in progress.

\begin{acknowledgments}

We thank Antoine Georges for helpful discussions.
This work was supported in part by the Air Force Office of Scientific Research
under AFOSR Award No.~FA9550-19-1-0405 and by the National Science Foundation
Cyberinfrastructure for Sustained Scientific Innovation program under Award
No.~OAC-1931321.  We acknowledge computing resources from Columbia University’s
Shared Research Computing Facility project, which is supported by NIH Research
Facility Improvement Grant 1G20RR030893-01, and associated funds from the New
York State Empire State Development, Division of Science Technology and
Innovation (NYSTAR) Contract C090171, both awarded April 15, 2010.  The Flatiron
Institute is a division of the Simons Foundation.

\end{acknowledgments}

\appendix

\section{Classical vibrational impurity solver}
\label{app:classical_imp}

For the nearest-neighbor interactions considered in the manuscript, only the
boundary atoms of cellular VDMFT are coupled to the bath. Therefore, the
interior atoms $\alpha$ and boundary atoms $\beta$ obey the coupled equations of
motion
\begin{subequations}
\label{eq:app_gle}
\begin{align}
\ddot{u}_{\alpha}(t) &= -\frac{dV_\mathrm{eff}}{d u_\alpha} \\
\ddot{u}_{\beta}(t) &= -\frac{dV_\mathrm{eff}}{d u_\beta}
    - \int_{0}^{t} d s \gamma_{\beta}\left(t-s\right) \dot{u}_{\beta}\left(s\right)
    + \xi_{\beta}(t),
\end{align}
\end{subequations}
where the random force satisfies detailed balance
$\langle \xi_\beta(t)\xi_\beta(s) \rangle = k_\mathrm{B}T \gamma_\beta(t-s)$.
As described, e.g., in Ref.~\onlinecite{Tuckerman1993}, $\xi_{\beta}(t)$ can be
expressed by a Fourier decomposition, the components of which are sampled from a
Gaussian distribution with a variance equal to the Fourier transform of the
memory kernel.  For each random sampling, a trajectory of length $T$ timesteps
can be calculated via explicit integration of the coupled integrodifferential
equations~(\ref{eq:app_gle}).  However, this introduces an $O(T)$ storage cost
and $O(T^2)$ computational cost.  To eliminate these costs, we simulate the
non-Markovian dynamics via Markovian dynamics in an extended
phase-space~\cite{Marchesoni1983,Ceriotti2009Langevin,Ceriotti2010}.

For each physical boundary coordinate, $u_\beta$, we add a set of $n$ auxiliary
momenta ${s}_i^\beta$, which have bilinear coupling to the physical momentum and
among themselves.  Decomposing the memory kernel in the form
\begin{equation}
\gamma_\beta(t) = \sum_{i=1}^{n}\sum_{j=1}^{n} a_i^\beta [e^{-\mathbf{A}^\beta t}]_{ij} a_j^\beta
    = [\bm{a}^\beta]^\mathrm{T} e^{-\mathbf{A}^\beta t} \bm{a}^\beta,
\end{equation} 
where $\mathbf{A}^\beta$ is a real, anti-symmetric matrix whose complex
eigenvalues have a positive real part, the equations of motion for the $(1+n)$
momenta are
\begin{subequations}
\begin{align}
\ddot{u}_\beta(t) &= -\frac{dV_\mathrm{eff}}{d u_\beta}
    - \sum_{j=1}^{n} {a}_{j}^\beta s_j^{\beta}(t), \\
\dot{s}_i^\beta(t) &= {a}_i^\beta \dot{u}_\beta(t) 
    - \sum_{j=1}^{n} {A}_{ij}^\beta s_j^{\beta}(t) + \sqrt{2k_\mathrm{B}T A_{ii}^\beta}\ {\zeta}_i^\beta(t),
\end{align}
\end{subequations}
where $\langle \zeta_i^\beta(t) \zeta_j^\beta(s)\rangle = \delta_{ij}\delta(t-s)$. 

A numerically convenient decomposition of the Fourier transform of the memory kernel,
\begin{equation}
\gamma(\omega) = \frac{2}{\pi}\sum_{i=1}^{m} \frac{\eta_i \gamma_i(\omega^2 + \omega_i^2 + \gamma_i^2)}
    {[(\omega+\omega_i)^2+\gamma_i^2][(\omega-\omega_i)^2+\gamma_i^2]},
\end{equation}
can be achieved with a simple structure of $\bm{a}^\beta$ and $\mathbf{A}^\beta$ in terms
of pairs of modes, 
\begin{subequations}
\begin{align}
a^\beta_{2i-1} = a^\beta_{2i} &= \sqrt{\eta_i/2\pi} \\
A^\beta_{2i-1,2i-1} = A^\beta_{2i,2i} &= \gamma_i \\
A^\beta_{2i-1,2i} = -A^\beta_{2i,2i-1} &= \omega_i
\end{align}
\end{subequations}
such that there are $n=2m$ auxiliary momenta.  In the results presented in the
manuscript, we use the above form to numerically fit the memory kernel with up
to $m=14$ modes (i.e., $n=28$ auxiliary momenta).

\section{Quantum vibrational impurity solver}
\label{app:quantum_imp}

First, we use grid techniques to numerically solve the Schr\"{o}dinger equation
for the anharmonic subsystem Hamiltonian $H_\mathrm{s}$ and keep the lowest
$N_\mathrm{s}$ eigenstates $\psi_a(u) = \langle u|\psi_a\rangle$,
depending on temperature. For the results presented in the manuscript, we kept
up to $N_\mathrm{s}=18$ eigenstates. The eigenstates are then transformed to a
discrete variable representation~\cite{Light1985}, 
$|d\rangle = \sum_a U_{ad} |\psi_a\rangle$ that diagonalizes the position
operator and thus makes the system-bath Hamiltonian diagonal,
\begin{subequations}
\begin{align}
H_\mathrm{s} &= \sum_{dd^\prime} |d\rangle H_{dd^\prime} \langle d^\prime| \\
H_\mathrm{sb} &= \sum_{d} |d\rangle u_{d}\langle d| \sum_m c_m x_m
\end{align}
\end{subequations}
where $u_{d} = \int du\ u\ |\phi_d(u)|^2$. The quantum correlation function is
then calculated as 
$\langle U(t) U(0) \rangle$ where $U = \sum_d |d\rangle u_d \langle d|$.

To simulate the quantum dynamics of an $N_\mathrm{s}$-level system linearly
coupled to a bath of harmonic oscillators, we use the hierarchical equations of
motion method~\cite{Tanimura1989,Ishizaki2005} as implemented in the pyrho
package developed in our group~\cite{pyrho}.  Similar to the classical impurity
solver, the spectral density of the bath is numerically fit to a sum of
underdamped Lorentzian modes~\cite{Liu2014},  
\begin{equation}
J(\omega) = \sum_{i=1}^{m} \frac{\eta_i \omega}
    {[(\omega+\omega_i)^2+\gamma_i^2][(\omega-\omega_i)^2+\gamma_i^2]};
\end{equation}
for the results in the manuscript, we used up to $m=8$ modes.  To simulate the
thermal correlation function, we first propagated the system and auxiliary
density matrices starting from a factorized initial condition, until reaching
equilibrium; this set of density matrices was then used as the initial condition
for the dynamics of the correlation function. All results were found to be
converged with the hierarchy truncated at level $L=4$ and $K=0$ Matsubara
frequencies.

\end{document}